\documentstyle[12pt]{article}
\begin{document}
\title{Complexified spinor fields in operator form}
\author{Slobodan Prvanovi\'c \\
{\it Institutes of Physics Belgrade,} \\
{\it University of Belgrade, Pregrevica 118, 11080 Belgrade, }\\
{\it Serbia}}
\date{}
\maketitle

\begin{abstract}
The formalism of quantum field theory in operator form, based on the anti self-adjoint operators of the imaginary coordinate and momentum and the self-adjoint 
operators of the real coordinate, momentum, energy and time, is used in considerations of the spinor fields and related topics. The unitary representation of the 
Lorentz boosts in the spin-orbital space is given. The operators that mirror-reflect the spin are introduced and then used in discussion of the spin parity. The 
conclusion that the spin is odd is used in the analyze of the parity violation in the cobalt-60 beta decay and it is found that the parity is not broken. The explanation why there are only right-handed antineutrinos and left-handed neutrinos is offered on the basis of appropriate treatment of the influence of spin on momentum. The inversion 
of time is treated within the framework of the operators of time and energy and this symmetry is represented in a way that respects the Schr\"odinger equation and 
this is done by the Wick rotations of involved operators and vectors of the complexified formalism. 

\end{abstract}

\section{Introduction}

Quantum field theory (QFT), it is a common opinion, is a very successful theory which still attracts much attention [1-20]. However, it has some unresolved problems, 
unsatisfactory solutions and misconceptions [21-24], some of which we have addressed in the previous article [25]. There, we have introduced complexified QFT in 
operator form, which was appropriate for scalar, electromagnetic, Proca and gravitational field. With the proposed formalism, we have tried to respect all important 
results and good features of QFT. Such an approach we shall keep in this article where we shall, as was said in the mentioned article, focus on the spinor fields.

In the center of the present investigation is the role of spin. There are many situations where the presence of spin is of crucial importance. For that reason, we 
shall focus on its relation to the momentum that characterize the modes of the considered fields and connected representation of the Lorentz boosts. Then, we shall 
thoroughly analyze representation of the parity, its (non)violation in the cobalt-60 beta decay and the question why there are only right-handed antineutrinos and 
left-handed neutrinos. Since the time inversion symmetry is closely related to the spatial inversion, we shall discuss it, as well, and this we are going to do 
by using our complexified formalism.  
 
Our approach is based on the self-adjoint operators of energy and time and the anti self-adjoint operators of coordinate and momenta. They allowed us to treat 
negative energies, that are unavoidable in QFT, in a consistent way. The fields within our approach are completely represented by operators, {\it i.e.}, they do 
not mix creation and annihilation operators with the functions. We shall give short review of our proposal regarding these operators in Section 2., and then we 
shall concentrate on the representations of symmetries. In Section 3. we shall discuss the effects of the spin influence on momentum, and then we shall propose the 
unitary representation of the Lorentz boosts. In Section 4. we shall discuss parity which would be treated as consisting of mirror reflections. In this section we 
shall introduce the operators that represent mirror reflections of the spin. Then, as will be shown, the spin will appear as the odd operator under 
parity and this will lead to conclusion that the parity is not broken in the case of the cobalt-60 beta decay. In Section 5. the complete representation of the time 
inversion will be given, and that we are going to do by employing our formalism where the energy, beside time, is represented by the operator and where there are 
anti self-adjoint operators of coordinate and momentum. Finally, some remarks will be given in the last section. 
 
\section{Basic definitions}

Within the standard treatment of quantum mechanics and QFT time is considered as the external parameter. As a consequence, there are some problems regarding 
interpretation of the Schr\"odinger equation as well as the statement that time inversion is represented via antilinear operator (this we shall discuss below). 
However, there are approaches where time and energy are treated on an equal footing with coordinate and momentum. We have done this in [25-32] while the operator of 
time is discussed in [33-35] and references therein. Our approach, being similar to the one given in [36], and references therein, and [37], starts with the 
introduction of the separate Hilbert space ${\cal H}_t$ where the operators of time $\hat t$ and energy $\hat s$ should act (it is understood that this is done in 
the same manner as it is done for each degree of freedom in the standard formulation of quantum mechanics). Then, the commutation relation for these operators is 
imposed in the same way as it is done for the operators of coordinate and momentum:
\begin{equation}
{1\over i\hbar} [\hat t , \hat s] = - \hat I.
\end{equation}
In complete analogy with the operators of coordinate and momentum, the operators of time and energy have the unbounded spectrum $(-\infty , +\infty)$ and the 
eigenvectors of $\hat t$ are $\vert t \rangle $ for every $t\in {\bf R}$. In time $\vert t \rangle$ representation, the operator of energy is differential operator 
$i \hbar {\partial \over \partial t}$, just like the operator of momentum is in the coordinate representation, and its eigenvectors $\vert E \rangle$ (in this 
representation) are $e^{{1\over i\hbar} E\cdot t}$ for every $E\in \bf R$. Obviously, within this formalism there are negative energies (as there are negative values 
of momentum in the standard formalism). However, the Schr\"odinger equation is seen here as a constraint in the overall Hilbert space that selects the states with 
non-negative energy for the usually used Hamiltonians. 

For the quantum system with three spatial degrees of freedom the complete space is ${\cal H}_x \otimes {\cal H}_y \otimes {\cal H}_z \otimes {\cal H}_t$. The 
operators of coordinates are $\hat r _x$, $\hat r _y$ and $\hat r _z$ and operators of momentum are $\hat k _x$, $\hat k _y$ and $\hat k _z$ (acting non trivially in 
appropriate spaces and being identical to those used in the standard formalism of quantum mechanics). The Hamiltonian is $H ({\hat {\vec r}}, {\hat {\vec k}})$ and 
the constraining equation:
\begin{equation}
 \hat s \vert \psi \rangle = H ({\hat {\vec r}}, {\hat {\vec k}}) \vert \psi \rangle  ,
\end{equation}
is nothing else but the Schr\"odinger equation. This becomes obvious after taking the coordinate-time $\vert x \rangle \otimes \vert y \rangle \otimes \vert z 
\rangle \otimes \vert t \rangle $ representation of this equation. Then one finds the familiar form of the Schr\"odinger equation:
\begin{equation}
 i \hbar {\partial \over \partial t} \psi (x, y, z, t) = H ({\vec r} , - i \hbar {\partial \over \partial {\vec r}} ) \psi (x, y, z, t) .
\end{equation}
Therefore, due to the non-negative spectra of the Hamiltonian $H ({\hat {\vec r}}, {\hat {\vec k}})$, the states with non-negative energies, that are only 
considered within the standard formalism of quantum mechanics, are being selected by the Schr\"odinger equation in a sense that only for such states the constraint (2) 
is satisfied. On the other side, as is well known, the negative energies occasionally do appear in quantum mechanics and are the common feature of QFT, so they have to 
be accepted and treated appropriately. 

Beside the standard self-adjoint operators of coordinate and momentum ${\hat {\vec r}} _{re} $ and ${\hat {\vec k}} _{re}$, which act in the rigged Hilbert spaces 
${\cal H}_{re, x} \otimes {\cal H}_{re, y} \otimes {\cal H}_{re, z} $ (the index $re$ stands in order to designate the real spectrum of these operators), the anti 
self-adjoint operators of coordinate and momentum ${\hat {\vec r}} _{im}$ and ${\hat {\vec k}} _{im}$ can be introduced. They act in the rigged Hilbert spaces 
${\cal H}_{im, x} \otimes {\cal H}_{im, y} \otimes {\cal H}_{im, z} $ (the index $im$ designates the imaginary spectrum of related operators). The spectral 
form of the imaginary coordinate $\hat r_{im, x}$, in the basis $\vert r_{im, x} \rangle$, where $x_{im}$ ranges over entire imaginary axis, is $\hat r_{im, x} = 
\int r_{im, x} \vert r_{im, x} \rangle \langle r_{im, x} \vert dr_{im, x}$, where $dr_{im, x}$ is the real measure (similarly for other two imaginary coordinates 
$\hat r_{im, y}$ and $\hat r_{im, z}$). In the basis $\vert r_{im, x} \rangle$, the anti self-adjoint operator of imaginary momentum $\hat k_{im, x}$ is represented by 
$-i \hbar {\partial \over \partial r_{im, x}}$ and its eigenvectors (in the same representation) are the "imaginary" plain waves $e^{-{1\over i \hbar} k_{im, x} \cdot 
r_{im, x}}$ with the imaginary eigenvalues. In parallel to the case of the real coordinate and momentum, the commutator of the imaginary ones is proportional to the 
$\hat I _{im}$, which is $\hat I _{im} = \int \vert r_{im, x} \rangle \langle r_{im, x} \vert dr_{im, x} $. (Similarly holds for other two imaginary momenta 
$\hat k_{im, y}$ and $\hat k_{im, z}$.)

If one uses Hamiltonian that is, for example, quadratic in the imaginary momentum, then its spectrum would be $(- \infty , 0]$, so the negative values of the operator 
of energy are connected to the Hamiltonian that is function of the imaginary coordinates and momenta. For the anti self-adjoint operators of coordinate and momentum 
the negative energies are as natural as the positive energies are for the self-adjoint coordinate and momentum. The situations (like tunneling [38]) where the 
imaginary values of momentum are unavoidable indicate that there are two worlds - one that is characterised by the real numbers and one that is characterised by the 
imaginary numbers. These two worlds together constitute the whole universe. The complete formalism has to contain both the real and the imaginary spatial coordinates and 
momenta and, consequently, both positive and negative values of energy should be treated as possible. The particular choice of the Hamiltonian would select, via 
the Schr\"odinger equation, which one would be realised by the quantum systems state. 

In contrast to the spatial coordinates and momenta, there are only self-adjoint operators of energy and time in our approach. However, as the components of quadri 
vectors of operators, one can introduce $\hat r ^0 _{re} = c_{re} \cdot \hat t$ and $\hat k ^0 _{re} = {\hat s \over c_{re}}$ for the real world and 
$\hat r ^0 _{im} = c_{im} \cdot \hat t$ and $\hat k ^0 _{im} = {\hat s \over c_{im}}$ for the imaginary world, where $c_{im}$ is the speed of light in the imaginary 
world, which is $c_{im} = i \cdot c $, ($c_{re} = c$). 

The unavoidable negative energy term, that appears in expression for the single component quantum field:
\begin{equation}
\hat \phi (\vec r , t) \sim
\int (\hat a_{\vec k} e^{-{1\over {i \hbar}} \vec k \cdot \vec r} e^{+{1\over {i \hbar}} E_{\vec k} \cdot t} + 
\hat a^{\dagger} _{\vec k} e^{+{1\over {i \hbar}} \vec k \cdot \vec r} e^{-{1\over {i \hbar}} E_{\vec k} \cdot t}) \cdot d\vec k ,
\end{equation}
can not be explained within the standard formalism of QFT. Beside this, there are some problems that are the consequence of the inconsistency of the formalism 
which mixes operators and functions, so we have proposed formalism based solely on the operators discussed above. More concretely, instead of the functions 
$e^{\pm {1\over {i \hbar}} \vec k \cdot \vec r}$ which represent the modes of the field, that are seen as the coordinate representation of the vector 
$\vert \vec k \rangle = \vert k_x \rangle \otimes \vert k_y \rangle \otimes \vert k_z \rangle $, the modes should be represented by the operators - dyads: 
$\vert \vec k \rangle \langle \vec k \vert$, and since the time and spatial coordinates should be treated equally, instead of $ e^{\pm {1\over {i \hbar}} 
E_{\vec k} \cdot t}$, that are seen as the time representation of the energy states, dyads $\vert E_{\vec k} \rangle \langle E_{\vec k} \vert$ should be used. 
Independence of degrees of freedom should be represented by introduction of the separate Hilbert spaces, which should be directly multiplied, and since both real 
and imaginary worlds have to be simultaneously represented, the mode of the quantum field characterised by ${\hat {\vec k}}_{re}$, $E _{\vec k _{re}}$, 
${\hat {\vec k}} _{im}$ and $E _{\vec k_{im}}$ is given by:
\begin{equation}
\vert \vec k_{re} \rangle \langle \vec k_{re} \vert \otimes \vert E_{\vec k_{re}} \rangle \langle E_{\vec k_{re}} 
\vert \otimes 
\vert \vec k_{im} \rangle \langle \vec k_{im} \vert \otimes \vert E_{\vec k_{im}} \rangle \langle E_{\vec k_{im}} \vert.
\end{equation}
(Since $E _{\vec k_{re}}$ and $E _{\vec k_{im}}$ are both real, it is possible to take common rigged Hilbert space where $ \vert E_{\vec k_{re} , \vec k_{im}} 
\rangle \langle E_{\vec k_{re} , \vec k_{im}} \vert $ represents the energy of the mode. However, we shall take two spaces and separate energy in a part that is 
connected to the real world and the one that is connected to the imaginary world.)  

Within the appropriate rigged Hilbert spaces where the above dyads appear, the operators of spatial coordinates ${\hat {\vec r}} _{re}$ and ${\hat {\vec r}} _{im}$, 
momenta ${\hat {\vec k}} _{re}$ and ${\hat {\vec k}} _{im}$ and operators of energy and time act and since the fields constitute of the normal modes and the amplitude, 
in order to address the spinor fields, two additional Hilbert spaces should be introduced. Within the "amplitude" rigged Hilbert space, the standard non-commuting 
operators $\hat q$ and $\hat p$ act, while in the spin space the well known Pauli matrices operate. After introduction of:
\begin{equation}
\hat q \otimes 
\vert \vec k_{re} \rangle \langle \vec k_{re} \vert \otimes \vert E_{\vec k_{re}} \rangle \langle E_{\vec k_{re}} 
\vert \otimes \vert \vec k_{im} \rangle \langle \vec k_{im} \vert \otimes \vert E_{\vec k_{im}} \rangle \langle 
E_{\vec k_{im}} \vert \otimes \hat I,
\end{equation}
\begin{equation}
\hat p \otimes 
\vert \vec k_{re} \rangle \langle \vec k_{re} \vert \otimes \vert E_{\vec k_{re}} \rangle \langle E_{\vec k_{re}} 
\vert \otimes \vert \vec k_{im} \rangle \langle \vec k_{im} \vert \otimes \vert E_{\vec k_{im}} \rangle \langle 
E_{\vec k_{im}} \vert \otimes \hat I ,
\end{equation}
the spinor field in the operator form is finaly defined as:
\begin{equation}
\hat \phi = \int \int \hat n \otimes 
\vert \vec k_{re} \rangle \langle \vec k_{re} \vert \otimes \vert E_{\vec k_{re}} \rangle \langle E_{\vec k_{re}} 
\vert \otimes \vert \vec k_{im} \rangle \langle \vec k_{im} \vert \otimes \vert E_{\vec k_{im}} \rangle \langle 
E_{\vec k_{im}} \vert \cdot d \vec k_{re} \cdot d \vec k_{im} \otimes \hat I.
\end{equation} 
The number operator $\hat n = \hat a^{\dagger} \cdot \hat a$ and the operators of creation and annihilation are constructed from $\hat q$ and $\hat p$ in the 
familiar way, of course. 

\section{Lorentz boosts}

The Dirac equation is usually solved in such a way that the solution for a rest frame is found and then this solution, as it is said, is boosted, so the solution 
for arbitrary momentum is acquired. In doing this, one actually uses what is called spinoral representation of the Lorentz boosts:
\begin{equation}
\left(\begin{array}{cc} 
 {\hat \omega} _{re, +} + 
{\hat k _{re, z} \over \sqrt {{{\hat {\vec k}} _{re}}^2}} \cdot {\hat \omega}_{re,-}
& 
({\hat k _{re, x} \over \sqrt {{{\hat {\vec k}} _{re}}^2}} + i \cdot 
{\hat k _{re, y} \over \sqrt {{{\hat {\vec k}} _{re}}^2}})
{\hat \omega}_{re,-}
\\ 
({\hat k _{re, x} \over \sqrt {{{\hat {\vec k}} _{re}}^2}} - i \cdot 
{\hat k _{re, y} \over \sqrt {{{\hat {\vec k}} _{re}}^2}})
{\hat \omega}_{re,-} 
& 
{\hat \omega}_{re, +} - 
{\hat k _{re, z} \over \sqrt {{{\hat {\vec k}} _{re}}^2}} \cdot {\hat \omega}_{re,-}
\end{array} \right) ,
\end{equation}
where
\begin{equation}
{\hat \omega}_{re, +} = {\sqrt {{{1\over \sqrt {{1-{{{\hat {\vec k}}_{re}^2 \over m^2 \cdot c_{re} ^2 +{\hat {\vec k}}_{re}^2}}} }}}+1 \over 2}} ,
\end{equation} 
\begin{equation}
{\hat \omega}_{re,-} = {\sqrt {{{1\over \sqrt {{1-{{{\hat {\vec k}}_{re}^2 \over m^2 \cdot c_{re} ^2 +{\hat {\vec k}}_{re}^2}}} }}}-1 \over 2}} ,
\end{equation} 
and where the standard Pauli matrices were used in the construction of this expression. The result of the application of this matrix on the momentum-spin vectors 
$\vert \vec k \rangle \otimes \vert s \rangle$ is that the magnitude of the spin becomes dependent on momentum because, as is the common opinion, this change is 
attached only to the spin part of the state and not to the orbital part, despite the fact that it is a change of the whole $\vert \vec k \rangle \otimes \vert s 
\rangle$ and the fact that boost is nothing else but the change of the (velocity and) momentum. Then, the values of momentum are present in both the orbital and 
spin space, which is formally incorrect and essentially wrong. Moreover, the application of the above matrix on $\vert \vec k \rangle \otimes \vert s \rangle$ is not 
related to the spinoral representation of genuine boost because for such a boost the common velocity (momentum) should be relativistically added to all states, while 
here each $\vert \vec k \rangle$ is individually "boosted" from the state with $\vec k = $0. However, this shows that the spin and momentum are interrelated. 

The standard approach to the representation of the Lorentz boosts in the spin space assumes that the boost, which is the change of the value of momentum, changes 
the magnitude of the spin, as well. That is, because of the change of the orbital part of the state of quantum system $\vert \vec k \rangle$, the spin state 
$\vert s \rangle$ is changed. Our approach to the representation of the Lorentz boosts is quite different. We start from the fact that the spin space has to be 
attached to the orbital Hilbert space in order to have the complete picture, so the unitary representation of the Lorentz boosts should not be searched for only 
within the spin space. Then, the mutual influence of the spin and momentum is treated in a way that is opposite to the standard one. That is, due to the presence of 
the spin, the momentum state is modified. The change of $\vert \vec k \rangle$, occurring in the orbital space, should be effectively equal to the change of 
$\vert s \rangle$ due to the boost as described within the standard treatment of the boosted spin state. Let us be more concrete. 

Regarding the orbital-spin observable ${\hat {\vec k}}_{re}\cdot {\hat {\vec s}}$, the state gained after the application of the operator matrix (9) on 
$\vert \vec k \rangle \otimes \vert s \rangle$ is indistinguishable from the state with unchanged magnitude of the spin, but with the appropriately modified momentum 
state. In the case when (9) acts on the state with spin up along $ z $ axis, the related state with everything related to the momentum being represented within the 
orbital part, is given by:
\begin{equation}
\left(\begin{array} {cc}
\Pi ^{\otimes} _{l}
\vert (\omega _{re, +} + {k _{re, z} \over \sqrt {{{{\vec k}} _{re}}^2}} \cdot \omega _{re,-}) k_{re, l} \rangle \\ 
e^{i \cdot \theta} \Pi ^{\otimes} _{l}
\vert \rho _{re, x, y} \cdot \omega _{re,-} \cdot k_{re, l} \rangle 
\end{array} \right) ,
\end{equation}
where $\omega _{re, +}$ and $ \omega _{re,-}$ are as their operator counterparts in (10) and (11), while $\rho _{re, x, y}$ and $\theta $ are modulus and argument of the 
complex number attached to the action of $\hat k _{re, x}$ and $\hat k _{re, y}$ in (9). The similar state would be appropriate for the spin down and the expressions 
will simplify in the case of helicity states. These states will be discussed in the next section and, before considering them, let us address the construction 
of the unitary operator that represents the boosts in the spin-orbital space. (For the sake of simplicity of expressions, at this place we shall discuss only the 
sector of the formalism that is connected to the real world.)

Starting from the expression for the change of momentum related to the transition from one coordinate system, where the momentum is ${\vec k}_{re}$, to the other 
coordinate system with relative velocity ${\vec v}_{re}$, the following expression can be introduced: 
$$
{\hat f}_{re, l} = 
{1\over {1- {{v}_{re, i} \cdot \sqrt {{{{{\hat {k}}_{re, i}^2 \over m^2 +{{\hat {k}}_{re, i}^2 \over c_{re} ^2}}}}}\over c_{re} ^2}}} \cdot 
$$
\begin{equation}
\cdot 
({\hat k_{re, l} \over \gamma _{{\vec v}_{re}}} - {m \cdot {\vec v}_{re, l} \cdot \gamma _{{\vec v}_{re}}
}+ {1\over c_{re}^2} \cdot {m \cdot \gamma _{{\vec v}_{re}}^2 \over 1+ \gamma _{{\vec v}_{re}}} \cdot (\sqrt {{{{{\hat {k}}_{re, j}^2 \over m^2 +
{{\hat {k}}_{re, j}^2 \over c_{re} ^2}}}}} \cdot {v}_{re, j}) \cdot v_{re, l}) - {\hat k}_{re, l} ,
\end{equation}
where 
\begin{equation}
\gamma _{{\vec v}_{re}}= {1\over \sqrt {1- {{{{\vec v}}}_{re}^2 \over c_{re}^2}}} ,
\end{equation}
and $i, j, l \in (x, y, z)$. Then, the operators: 
\begin{equation}
e^{{-{1\over i \cdot \hbar} {\hat r}_{re, l} \cdot {\hat f}_{re, l}} },
\end{equation}
are the unitary operators representing boosts in the spin-orbital space. (It is understood that in (15) symmetrised product is used.) These operators transform 
$\vert k_{re, x} \rangle \otimes \vert k_{re, y} \rangle \otimes \vert k_{re, z} \rangle $, that describes the state of the momentum in one coordinate system, to the boosted state as it is seen in the coordinate system with relative velocity ${\vec v}_{re}$ and they should be applied to the above discussed "spin modified" momentum states if such states have to be boosted. 

Boosts in the imaginary world can be treated in a similar way, with some modifications of the above expressions related to the substitution of ${\hat {\vec k}}_{re}$, 
${\vec k}_{re}$ and $c_{re}$ by the corresponding imaginary counterparts.

\section{Parity}

Parity is transformation of spatial coordinates such that all three of them change the sign. Under parity, the components of momentum change sign, as well, so both 
basic observables are odd, due to which their commutation relations are invariant. The parity can be seen as the symmetry transformation that consists of three 
mirror reflections, each of which changes the sign of just one spatial coordinate. In difference to the boosts, parity does not change the magnitude of 
${\vec r}_{re}$ or ${\vec k}_{re}$, so it might be said that parity is just the mere relabeling that occurs when there is transition from one coordinate system to 
the other, the inverted one. 

Since the parity deals with the orientation of coordinate axes, it is important to find out how this transformation affects the spin. Usually it is said that the spin is even under parity transformation because the angular momentum is even. In some sense, it is true that the spin and 
angular momentum are observables of the same kind regarding their nature. But, spin is the basic observable, while the angular momentum is the function of the basic 
orbital observables, so they are quite different and to rest the argument about the character of spin under parity just on its analogy to the angular momentum might 
be incorrect. Actually, the assumption that the spin is even under parity is wrong. This leads to the conclusion that parity is not broken in the cobalt nucleus decay, 
{\it i.e.}, there is no violation of parity symmetry in weak interaction. Let us be more concrete on this.

As it is said, parity might be seen as consisting of mirror reflections. Mirror reflection of the spatial axis that is orthogonal to the mirror is equivalent to the 
rotation of this axis by $\pi $ around some axis laying in the plane of the mirror. Then, in order to find out how the mirror reflection is represented in spinor 
space, one can apply the representation of rotation group. For example, the mirror reflection of the $z$ axis, when the mirror is in the $x-y$ plane, is equivalent 
to the rotation by $\pi $ around $x$ axis and then the state $\vert z_{re} \rangle$ is transferred to $\vert - z_{re} \rangle$. In the spin space rotations are 
represented by:
\begin{equation}
e^{i \cdot {\alpha \over 2} \cdot ({\vec n} \cdot {\hat {\vec \sigma}})} = \hat I \cdot cos{\alpha \over 2} + i \cdot ({\vec n} \cdot {\hat {\vec \sigma}}) 
sin{\alpha \over 2}.
\end{equation}
So, if the state of the spin in direction of the $z$ axis was $\vert \uparrow \rangle $, after the application of the matrix that represents considered rotation, 
which is:
\begin{equation}
\left(\begin{array}{cc} 
0
& 
i
\\ 
i
& 
0
\end{array} \right) ,
\end{equation} 
the state of the spin becomes $i \cdot \vert \downarrow \rangle $. Essentially the same result would follow for representation of spatial rotations by $\pi $ around $y$ 
axis. This leads to the conclusion that mirror reflection of spin up state is spin down state and {\it vice versa}. Therefore, one can introduce three reflectors, 
{\it i.e.}, the operators that represent mirror reflections in the spin space. For the mirrors in $x-y$, $y-z$ and $z-x$ plane, the reflectors are: 
\begin{equation}
\hat R _{x-y} = {1 \over {\sqrt {{a^z _x}^2 + {a^z _y}^2}}} (a^z _x \hat \sigma _x + a^z _y \hat \sigma _y),
\end{equation}
\begin{equation}
\hat R _{y-z} = {1 \over {\sqrt {{a^x _y}^2 + {a^x _z}^2}}} (a^x _y \hat \sigma _y + a^x _z \hat \sigma _z),
\end{equation}
\begin{equation}
\hat R _{z-x} = {1 \over {\sqrt {{a^y _z}^2 + {a^y _x}^2}}} (a^y _z \hat \sigma _z + a^y _x \hat \sigma _x) ,
\end{equation}
respectively, where all involved coefficients $a$ and $b$ are real. These operators are the Hermitian and unitary.

Obviously, as the components of the spin operator do not commute, the reflectors do not commute, as well. That is, the reflectors are non-compatible, 
so only one might be applied. The orbital and the inner (spin) spaces differ in that there are three Hilbert spaces, one for each spatial direction, in the case of 
the orbital space, while there is only one inner space which is, so to say, common arena for all three directions. Which one of (18-20) is appropriate for the case 
under consideration is determined by the actual preparation of the state. If, for example, the spin is prepared to be along $z$ axis, then ${\hat R}_{x, y}$ is the 
spinoral counterpart of the parity transformation. 

Therefore, when the parity transformation is inverting states in all three spatial spaces ${\cal H}_{re, x} \otimes {\cal H}_{re, y} \otimes {\cal H}_{re, z} $ it is 
accompanied with just one inversion within the spin space. No matter what is the spin state it would be inverted under parity transformation to the state that is the 
opposite one along the same direction. From this it follows that the spin is odd under parity. Namely, each component of the spin is the operator which changes sign 
after application of the adequate reflector (applied from the left and from the right side of the operator of spin). Put in another words, parity in orbital space 
consists of three separate and specified reflections while, in the spin space, it consists of just one among three possible mirror reflections, and that one 
becomes specified when direction of the spin state preparation/measurement is specified.  

Ever since the Wu experiment [39], it is a common opinion that there is parity violation in the beta decay of cobalt-60 nucleus, which is related to the weak interaction. 
However, assumption that lead to such conclusion was that the spin is even under parity transformation and this assumption, as is clear from the above discussion, is 
wrong. 

In the Wu experiment, the uniform magnetic field (in direction of $z$ axis) aligned the nuclear spins of cobalt-60, and then they were cooled down to almost absolute 
zero so that the thermal motions of the atoms could not ruin the spin alignment. As a result of the beta decay of cobalt-60, almost all of the emerging electrons had 
momentum in the negative direction of the $z$ axis, while the antineutrinos had momentum in the positive direction of the $z$ axis, {\it i.e.}, the antineutrinos had 
positive helicity. Due to this asymmetry in the distribution, it was concluded that the parity is not conserved in beta decay of cobalt-60.  

For the observer attached to the inverted coordinate system, {\it i.e.}, the coordinate system with the $z$ axis pointing in 
the opposite direction, the mentioned antineutrinos would be characterized as left-handed (negative helicity) because, as it is commonly assumed, the spin is even, 
so its projection on a direction of the momentum (which is odd under parity) would be negative. But, left-handed antineutrinos have not been observed ever in any 
experiment. 

The spin is odd, as is the momentum, so whether it is looked from the original or the inverted (mirror-reflected) coordinate system, helicity is the same. That is, 
the antineutrinos would be right-handed no matter of which coordinate system is chosen. This means that the parity symmetry is not broken. However, important 
question is why the antineutrinos are always right-handed and neutrinos left-handed and the explanation comes from the considerations given in the previous section. 
 
If $x$ and $y$ component of the real momentum are equal to zero, according to (12), the state of positive helicity is:
\begin{equation}
\vert 0 \rangle \otimes \vert 0 \rangle \otimes \vert (\omega _{re, +} + \omega _{re,-}) k_{re, l} \rangle \otimes \vert \uparrow \rangle ,
\end{equation}
while the state of negative helicity is:
\begin{equation}
\vert 0 \rangle \otimes \vert 0 \rangle \otimes \vert (\omega _{re, +} - \omega _{re,-}) k_{re, l} \rangle \otimes \vert \downarrow \rangle.
\end{equation}
Since the antineutrinos have (almost) vanishing mass, it holds:
\begin{equation}
\omega _{re, +} \approx \omega _{re,-}.
\end{equation}
Therefore, the $z$ component of the real momentum, when the state of antineutrino is the one with negative helicity, becomes (almost) equal to zero. The antineutrino, 
in general, has to cary some significant momentum, so the only way in which this can be realised is if it is in a state with positive helicity. This explains way 
the antineutrinos are always right-handed. On the other side, so to say, the neutrinos are characterized with the other chirality. More precisely, instead of (9), the 
matrix:
\begin{equation}
\left(\begin{array}{cc} 
 {\hat \omega} _{re, +} - 
{\hat k _{re, z} \over \sqrt {{{\hat {\vec k}} _{re}}^2}} \cdot {\hat \omega}_{re,-}
& 
-({\hat k _{re, x} \over \sqrt {{{\hat {\vec k}} _{re}}^2}} + i \cdot 
{\hat k _{re, y} \over \sqrt {{{\hat {\vec k}} _{re}}^2}})
{\hat \omega}_{re,-}
\\ 
-({\hat k _{re, x} \over \sqrt {{{\hat {\vec k}} _{re}}^2}} - i \cdot 
{\hat k _{re, y} \over \sqrt {{{\hat {\vec k}} _{re}}^2}})
{\hat \omega}_{re,-} 
& 
{\hat \omega}_{re, +} + 
{\hat k _{re, z} \over \sqrt {{{\hat {\vec k}} _{re}}^2}} \cdot {\hat \omega}_{re,-}
\end{array} \right).
\end{equation}
should be used in order to find how the presence of spin influences the momentum, and then the $z$ component of the real momentum of the states with positive 
helicity would be suppressed. Hence, the neutrinos with non-vanishing momentum can be only left-handed.  

\section{Time inversion}

Within the standard formulation of quantum mechanics, where time is treated as a parameter, the time inversion, denoted by $T$, is seen (just) as the transformation 
$t \rightarrow -t$. This makes the first difference between time and spatial inversions. Namely, for the former the energy, which should be the conjugated observable 
to time, is not changed, while for the later both conjugated observables change signs under inversion. Then, after Wigner, by assuming that the time independent 
Hamiltonian is not affected by time inversion, and due to the mentioned characterization of time as a parameter, the invariance of the Schr\"odinger equation 
demands $T$ to be an antilinear operator. That is, if the transformed state satisfies the Schr\"odinger equation with the inverted time, then in simplified notation:
\begin{equation}
 i \hbar {\partial (T \psi (t)) \over \partial (- t)} = \hat H T \psi (t) .
\end{equation}
So, by applying $T^{-1}$ from the left, there is:
\begin{equation}
 T^{-1} (-i) T \hbar {\partial \psi (t) \over \partial t} = T^{-1} \hat H T \psi (t)  ,
\end{equation}
and since $T^{-1} \hat H T = \hat H $, it follows that $T^{-1} (-i) T =i$. 

Essential in this reasoning was that the time has not been represented by an operator, but appeared as a parameter. If time is taken to be the operator, then there 
is the conjugated operator of energy, which (in $\vert t \rangle$ representation) is differential operator that appears on the LHS of (25) and which was introduced in 
Section 2. In this case, when $T$ is extracted from $(T \psi (t))$ on the LHS of (25), and then applied to the left, it would not pass through partial derivative with 
respect to $t$. There would be $ T^{-1}(- i \hbar {\partial \psi (t) \over \partial t} )T$, which is equal to the representation of $\hat s$. So, there would be 
no demands to introduce anti linearity into play in order to keep the Schr\"odinger equation satisfied. 

Within the approach we are proposing, time and energy appear as operators. On the other side, time and energy should be on an equal footing with coordinate and 
momentum. On the ground of inversions, this would mean that not just that the time should be transformed by time inversion, but the operator of energy should 
change the sign, too, and that in order to behave as the momentum under parity transformation. The reason for demanding this is obvious - time inversion is 
nothing else but the mirror reflection of the temporal component of the quadri vector of coordinate. As the mirror reflections of the spatial components of the quadri 
vector of coordinate are followed by the change of sign of the corresponding components of the quadri vector of momentum, time inversion has to change the sign of 
energy (energy divided by the speed of light is one of the four components of the quadri vector of momentum). In this way, when both $\hat t$ and $\hat s$ change 
signs under time inversion, their commutation relations remain unchanged. 

Hence, in complete analogy to spatial inversion, time inversion is:
\begin{equation}
(\hat t , \hat s) \rightarrow (- \hat t , - \hat s) ,
\end{equation}
or regarding vectors:
\begin{equation}
(\vert t_0 \rangle , \vert E_0 \rangle) \rightarrow (\vert - t_0 \rangle , \vert - E_0 \rangle) .
\end{equation} 

The non-negative energies are characteristic for the real world, while the non-positive energies characterize the imaginary world. Then, the change of sign of the 
energy due to the time inversion means that the roles of the real and imaginary sector in the formalism are interchanged. This means that to the time inversion, 
happening in the temporal Hilbert space, in the spatial Hilbert spaces is attached the Wick rotation:
\begin{equation}
({\hat {\vec r}}_{re} \otimes \hat I , {\hat {\vec k}}_{re} \otimes \hat I) \rightarrow ({\hat {\vec r}}_{im} \otimes \hat I , {\hat {\vec k}}_{im} \otimes \hat I) ,
\end{equation}
\begin{equation}
(\hat I \otimes {\hat {\vec r}}_{im} , \hat I \otimes {\hat {\vec k}}_{im}) \rightarrow (\hat I \otimes {\hat {\vec r}}_{re} , \hat I \otimes {\hat {\vec k}}_{re}) ,
\end{equation}
which should be accompanied with:
\begin{equation}
(\vert {\vec r}^a _{re} \rangle , \vert {\vec k}^a _{re} \rangle) \rightarrow (\vert {\vec r}^a _{im} \rangle , \vert {\vec k}^a _{im} \rangle) ,
\end{equation}
\begin{equation}
(\vert {\vec r}^b _{im} \rangle , \vert {\vec k}^b _{im} \rangle) \rightarrow (\vert {\vec r}^b _{re} \rangle , \vert {\vec k}^b _{re} \rangle) ,
\end{equation}
and the interchange of $c_{re}$ and $c_{im}$, where, for instance, ${\vec k}^a _{re} = {\vec k}^a$, while ${\vec k}^a _{im} = i \cdot {\vec k}^a$, and 
${\vec k}^b _{re} = {\vec k}^b$, while ${\vec k}^b _{im} = i \cdot {\vec k}^b$. 

For the Hamiltonian of Weyl spinors let us propose the following function of the basic operators:
\begin{equation}
+{\sqrt {({({\hat {\vec k} _{re}} \otimes \hat I )\cdot {\hat {\vec \sigma}} \cdot c_{re}})^2 + m^2 c_{re}^4}} - 
{\sqrt {({(\hat I \otimes {\hat {\vec k} _{im}} )\cdot {\hat {\vec \sigma}} \cdot c_{im}})^2 + m^2 c_{im}^4}} ,
\end{equation}
where ${\hat {\vec \sigma}}$ are the Pauli matrices. After the Wick rotation (29-30), the Hamiltonian is the same function of transformed operators:
\begin{equation}
+{\sqrt {({(\hat I \otimes {\hat {\vec k} _{re}} )\cdot {\hat {\vec \sigma}} \cdot c_{re}})^2 + m^2 c_{re}^4}} -
{\sqrt {({({\hat {\vec k} _{im}} \otimes \hat I ) \cdot {\hat {\vec \sigma}} \cdot c_{im}})^2 + m^2 c_{im}^4}}.
\end{equation}
The action of this Hamiltonian on $\vert {\vec k}^a _{im} \rangle \otimes \vert {\vec k}^b _{re} \rangle$ produces the same value, but the opposite sign, as when (33) 
acts on $\vert {\vec k}^a _{re} \rangle \otimes \vert {\vec k}^b _{im} \rangle$. Consequently, the validity of the Schr\"odinger equation after time inversion is 
maintained.

\section{Concluding remarks}

In this article we have continued development of the complexified quantum field theory in operator form, being concentrated on the spinor fields. Since the most 
important concerning spinors, from our point of view, is the representation of parity and related topics, we have discussed effects of the spin on momentum and 
representations of boosts, mirror reflections and time inversion.

The effects of the influence of the spin states on the momentum states we have incorporated within the momentum states. The spin modified momentum states that we have 
found are mutually orthogonal and normalized to $\delta (0)$, just like it is the case for the standard momentum eigenvectors. On the other side, these states allowed 
us to find the explanation why there are only left-handed neutrinos and right-handed antineutrinos. Moreover, we have found the unitary representation of the Lorentz 
boosts within the spin-orbital space, which is the space that has to be used, not just the inner space, when the unitary representation of the Lorentz boosts is 
considered.  

The parity transformation, which is seen as constituting of three mirror reflections of just one of all three axes, is accompanied by one mirror reflection within 
the spin space. We have introduced three non compatible operators - reflectors, that formalize these spin reflections. Then, after finding that the spin is odd 
under parity, we have shown that the parity is not broken in the cobalt-60 beta decay. The observer attached to the inverted coordinate system would see exactly 
the same phenomenon, which is that all antineutrinos are right handed, as would see the observer attached to the original coordinate system. This means that the parity 
is not violated in the weak interaction that governs the cobalt-60 beta decay.

The time inversion is completely analogous to the spatial inversion. It is seen as the symmetry that is just the mere relabeling of the time and energy. Within the 
cmplexified theory, the time inversion is connected to the unnoticeable interchange of the real and imaginary sectors of the formalism. Nothing essentially important 
happens if one puts the minus sign in front of the digits of the clock and the apparatus measuring energy, which is the practical realization of the time inversion 
that is considered theoretically. This is exactly the same sort of physically irrelevant changes as is the change when the observer, so to say, turns around in order 
to practically realise the spatial inversion. So, as symmetries, these inversions are truly just relabelings, in deep contrast to the boosts. 

We have proposed the Hamiltonian for the Weyl spinors. In the previous article we have given the Hamiltonian for bosons, so by combining them and by introducing 
interaction terms in the Hamiltonian, one can construct the Hamiltonians for the whole variety of situations. The creation and annihilation operators acting in the 
so called amplitude spaces offer the possibility to construct appropriate interaction terms. In the case of massless fermions, the proposed Hamiltonian for the free 
field simplifies. But, the attention has to be paid to the chirality of the considered systems, whether $+{\hat {\vec \sigma}}$ or $-{\hat {\vec \sigma}}$ is used, 
which is where the neutrinos and antineutrinos differ, and that in order to avoid inappropriate sign of the energy at some places within the formalism.

\section{Acknowledgment}

The author acknowledges funding provided by the Institute of Physics Belgrade, through the grant by the Ministry of Education, Science and Technological Development 
of the Republic of Serbia.

\end{document}